\documentclass[prl,reprint,twocolumn,superscriptaddress,noshowpacs,notitlepage,longbibliography]{revtex4-1}%
\usepackage{graphicx,bm,times}
\usepackage{amsmath}
\usepackage{amsfonts}
\usepackage{amssymb}
\usepackage{color}
\usepackage{hyperref}
\hypersetup{
	colorlinks = true,
	allcolors = {blue}
}

\begin{document}

\title{Probing the reconstructed Fermi surface of antiferromagnetic BaFe$_2$As$_2$ in one domain}

\author{Matthew D. Watson}
\email[corresponding author:]{mdw5@st-andrews.ac.uk}
\affiliation{Diamond Light Source, Harwell Campus, Didcot, OX11 0DE, United Kingdom}
\affiliation{School of Physics and Astronomy, University of St. Andrews, St. Andrews KY16 9SS, United Kingdom}

\author{Pavel Dudin}
\affiliation{Diamond Light Source, Harwell Campus, Didcot, OX11 0DE, United Kingdom}

\author{Luke C. Rhodes}
\affiliation{Diamond Light Source, Harwell Campus, Didcot, OX11 0DE, United Kingdom}
\affiliation{Department of Physics, Royal Holloway, University of London, Egham, Surrey, TW20 0EX, United Kingdom}

	\author{Daniil V. Evtushinsky}
\affiliation{Institute of Physics, Ecole Polytechnique Federale Lausanne,CH-1015 Lausanne, Switzerland}

	\author{Hideaki Iwasawa}
\affiliation{Diamond Light Source, Harwell Campus, Didcot, OX11 0DE, United Kingdom}
\affiliation{Graduate School of Science, Hiroshima University, Higashi-Hiroshima, Hiroshima 739-8526, Japan}

\author{Saicharan Aswartham}
\affiliation{Leibniz Institute for Solid State and Materials Research, 01171 Dresden, Germany}

\author{Sabine Wurmehl}
\affiliation{Leibniz Institute for Solid State and Materials Research, 01171 Dresden, Germany}

\author{Bernd B\"{u}chner}
\affiliation{Leibniz Institute for Solid State and Materials Research, 01171 Dresden, Germany}
\affiliation{Institut für Festkörperphysik, Technische Universität Dresden, D-01171 Dresden, Germany}

\author{Moritz Hoesch}
\affiliation{Diamond Light Source, Harwell Campus, Didcot, OX11 0DE, United Kingdom}
\affiliation{DESY Photon Science, Deutsches Elektronen-Synchrotron, Hamburg, Germany}

\author{Timur K. Kim}
\email[corresponding author:]{timur.kim@diamond.ac.uk}
\affiliation{Diamond Light Source, Harwell Campus, Didcot, OX11 0DE, United Kingdom}

\begin{abstract}

We revisit the electronic structure of BaFe$_2$As$_2$, the archetypal parent compound of the Fe-based superconductors, using angle-resolved photoemission spectroscopy (ARPES). Our high-resolution measurements of samples ``detwinned" by the application of a mechanical strain reveal a highly anisotropic 3D Fermi surface in the low temperature magnetic phase. By comparison of the observed dispersions with \textit{ab-initio} calculations, 
we argue that overall it is magnetism, rather than orbital ordering, which is the dominant effect, reconstructing the electronic structure across the Fe $3d$ bandwidth. Finally, we measure band dispersions directly from within one domain without applying strain to the sample, by using the sub-micron focused beam spot of a nano-ARPES instrument.      

\end{abstract}
%\date{\today}
\maketitle

%%%%%%%%%%%%%%%%%%%%%%%%%%%%%%%%%%%%%%%%%%%%%%%%

% INTRO
\begin{figure*}[t]
	\centering
	\includegraphics[width=\linewidth]{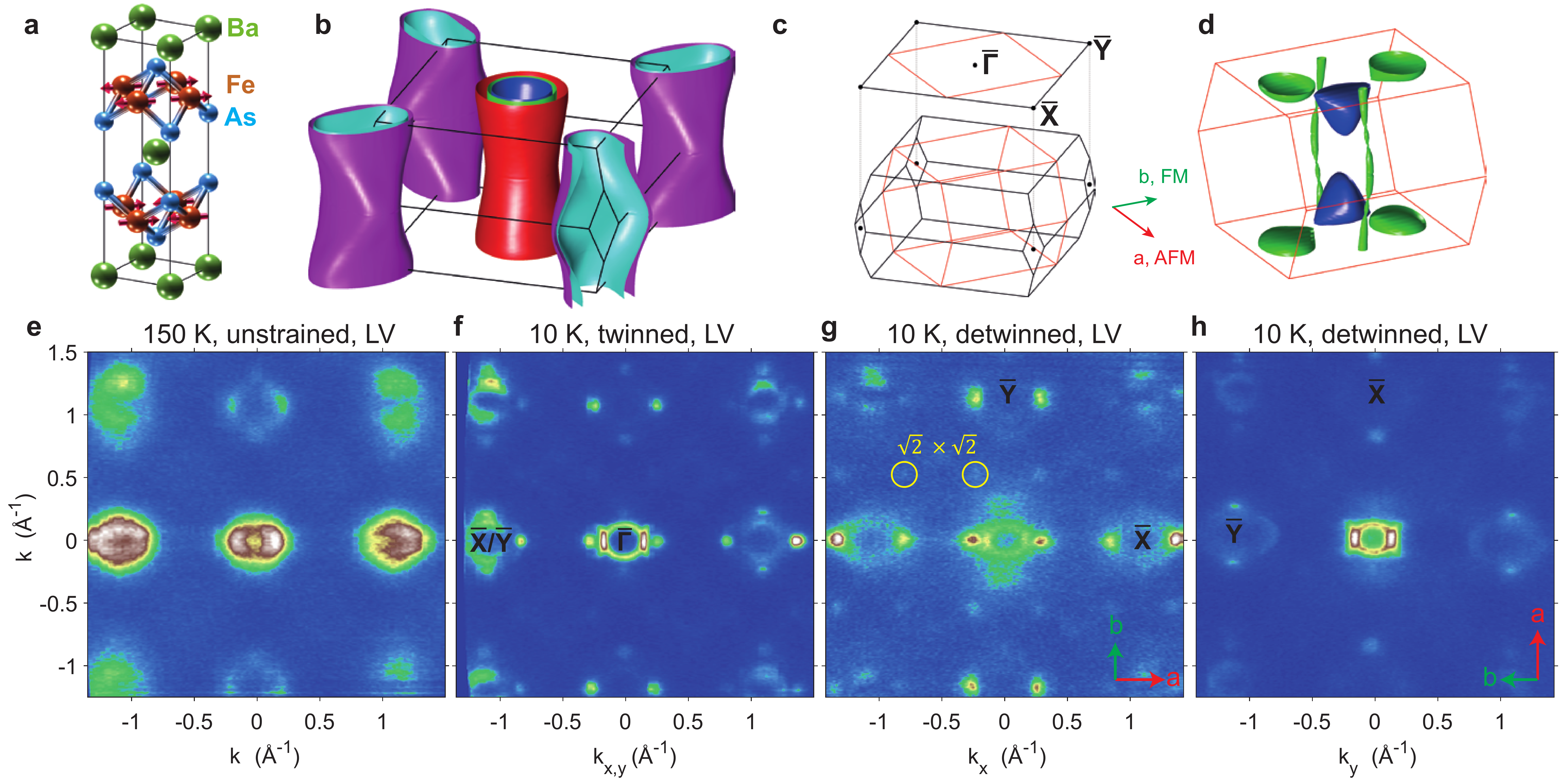}
	\caption{\textbf{Reconstructed Fermi surface in the AFM phase.} (a) Crystal structure of BaFe$_2$As$_2$, in the high-temperature tetragonal $I4/mmm$ space group. Arrows represent the orientation of magnetic moments on the Fe sites below $T_N$ (fully described in the orthorhombic space group $Cccm$, not shown). (b) Calculated non-magnetic Fermi surface of BaFe$_2$As$_2$ in the tetragonal phase. (c) Geometry of the reconstruction of the Brillouin zone in the stripe AFM phase, and the definitions of the labels used in this paper, in an effective surface Brillouin zone. (d) Calculated Fermi surface in the reconstructed phase (blue=hole-like, green=electron-like). e-h) Fermi surface maps obtained using 100 eV photons in Linear Vertical (LV) polarisation. Note that panel (h) is obtained on the same sample and with the same measurement geometry as (g), only the sample is rotated 90$^\circ$ azimuthally. Panels (e,f) were obtained on a separate unstrained reference sample.}
	\label{fig:fig1}
\end{figure*}

Unconventional superconductivity in the Fe-based systems emerges in proximity to antiferromagnetic (AFM) and ``nematic" phases, which are both characterised by pronounced magnetic and electronic anisotropies. Since the fluctuations associated with these  phases are likely to be important for the superconducting pairing, the ordered phases of the parent compounds, of which BaFe$_2$As$_2$ is the archetype, have been the subject of intense investigations. While in BaFe$_2$As$_2$ the orthorhombic structural (nematic) and magnetic transitions occur simultaneously at $T_s=T_N=$~137~K, the decoupling of the two transition temperatures in Ba(Fe$_{1-x}$Co$_{x}$)$_2$As$_2$ \cite{Nandi2010PRL}, NaFeAs \cite{Wright2012PRB}, and FeSe \cite{BoehmerKreiselReview2017} has caused much debate about the relative importance of magnetic and orbital degrees of freedom \cite{Fernandes2014,Yi2017review}. Some theoretical studies have interpreted the nematic (non-magnetic orthorhombic) phase as still being essentially magnetically-driven \cite{Fernandes2012PRB,Fernandes2014}, with a critical role played by significant biquadratic exchange interactions \cite{Wysocki2011NPhys,Glasbrenner2015NPhys}. However, another view is that the presence of a non-magnetic symmetry-breaking transition shows that orbital degrees of freedom need to be treated on at least an equal footing \cite{Baek2014,Yamakawa2016PRX,Yi2017review,Kissikov2018}. It is therefore important to experimentally probe the ground state of the parent compounds, but a recurring experimental challenge is that when fourfold symmetry is spontaneously broken at $T_s$, the samples naturally form orthorhombic twin domains, restoring macroscopic fourfold symmetry and masking the underlying anisotropies. To address this, it was found that samples could be effectively ``detwinned" by the application of a mechanical strain along the Fe-Fe direction \cite{Tanatar2010PRB}. Experiments performed on such ``detwinned" samples have yielded several important breakthroughs, in particular revealing pronounced intrinsic in-plane electronic anisotropies in measurements of resistivity and optical spectroscopy \cite{Chu2010Science,Fisher2011review}, phonon frequencies split by magnetic interactions \cite{Baum2018PRB}, and strongly anisotropic magnetic excitations in inelastic neutron scattering \cite{Lu2018PRL_Dai}. 

In the case of angle-resolved photoemission spectroscopy (ARPES), the beam spot of conventional systems is typically much larger than the size of the structural domains, and thus a superposition of the electronic structures from 90$^\circ$-rotated domains is normally observed experimentally - doubling the number of observed bands and complicating the interpretation. While early measurements of such twinned samples suggested a reconstruction of the electronic structure below $T_N$ \cite{Zabolotnyy2009,Richard2010PRL,Shimojima2010PRL,deJong2010EPL,Wang2013PRBDessau}, a more detailed picture started to emerge with the first ARPES reports on detwinned samples of NaFeAs \cite{Yi2012NJP,Zhang2012PRB_NaFeAs_Feng} and BaFe$_2$As$_2$ \cite{Kim2011PRB_Kwon,Yi2011PNAS,Sonobe2018}. Those results were influential in guiding the ideas of electronically-driven nematicity \cite{Fernandes2014,Yi2017review}, and in particular it was claimed \cite{Yi2011PNAS,Yi2012NJP,Zhang2012PRB_NaFeAs_Feng,Pfau2019PRB} that the detwinned results provided evidence for a large 50-80 meV energy scale associated with the splitting of bands with $d_{xz}$ and $d_{yz}$ orbital character which are degenerate in the tetragonal phase. However, recent studies of NaFeAs \cite{Watson2018PRB_NaFeAs}, FeSe \cite{Watson2016PRB,Watson2017NJP,Fedorov2016SciRep} and BaFe$_2$As$_2$ \cite{Fedorov2018_arxiv} have questioned the existence of any large nematic energy scale. Moreover, the ARPES measurements on BaFe$_2$As$_2$ in the literature were never fully reconciled with the Fermi surface determined by quantum oscillations \cite{Terashima2013PRL}. Meanwhile, alongside the general improvements in state-of-the-art ARPES systems and perfection of crystal growth techniques, a new opportunity is presented by the technical development of nano-ARPES, where the beam can be focused to sub-micron areas, allowing the possibility of directly measuring ARPES in one domain - doing away with the necessity for straining the sample entirely. There is therefore a strong motivation to revisit the electronic structure of BaFe$_2$As$_2$ in one domain with ARPES. 

In this paper we use high-resolution ARPES and nano-ARPES to find new insights into the electronic structure of the archetype parent compound BaFe$_2$As$_2$. We first present high-resolution measurements of a sample which was tuned from being twinned into an almost fully-detwinned sample by the application of strain \textit{in-situ}. The electronic structure in one domain  is found to match the 3D Fermi surface as determined by quantum oscillations. We reproduce a notable separation of bands along $\mathrm{\bar{\Gamma}-\bar{X}}$ and $\mathrm{\bar{\Gamma}-\bar{Y}}$ directions, previously interpreted in terms of a $d_{xz}$ - $d_{yz}$ orbital splitting. However, we argue that this energy scale should not be considered as a proxy for the energy scale of orbital ordering. Rather, our conclusion is that the electronic structure is strongly reconstructed across the whole bandwidth by the stripe AFM order. In addition, we used a nano-ARPES end station with a beam spot of less than 1 $\mu$m to directly measure ARPES spectra in one domain without requiring the application of any strain. These results show consistence with the high-resolution strained measurements, and allow us to use anisotropic features in the electronic structure to map the extensive stripe-like structural domains in real space.

\subsection{The Fermi surface: theory and experiment}

One of the most significant similarities between the cuprates and the Fe-based superconductors is that in both cases the parent compounds are antiferromagnetic, but a notable difference is that the parent compounds of Fe-based superconductors remain semimetallic \cite{Mazin2010} below $T_N$, albeit with a significantly reduced carrier density. A broad overview of the effects of this AFM order on the electronic structure of BaFe$_2$As$_2$, from the Density Functional Theory (DFT) perspective, is presented in Fig.~\ref{fig:fig1}(a-d). When enforcing a non-magnetic solution, the calculation yields a compensated multiband Fermi surface in Fig.~\ref{fig:fig1}(b), typical of Fe-based superconductors. This consists of three hole-like pockets at the centre of the Brillouin zone, and two electron-like pockets at the zone corner. In the ground state, however, the ($\pi$,0,$\pi$) stripe AFM order of BaFe$_2$As$_2$ (Fig.~\ref{fig:fig1}(a)) doubles the size of the unit cell and maps the electron dispersions at the Brillouin zone corner back to the zone centre, where they hybridise strongly with the hole dispersions. However, the Fermi surface is not fully gapped, and the resulting Fermi surface in the magnetic phase can be qualitatively understood by considering three concepts. First, the size and shapes and of the pockets are not perfectly matched in the normal phase, particularly if one takes into account their warping along $k_z$, so the backfolding and hybridisation leaves behind small, typically 3D, pockets. Second, it has been shown that certain band hybridisations in the stripe AFM phase are forbidden exactly on the high-symmetry axes, enforcing a so-called ``nodal SDW" \cite{Ran2009PRB}; this leads to an expectation of tiny pockets with Dirac-like band dispersions localised on the high-symmetry axes \cite{Richard2010PRL}. Finally, charge compensation ensures that both hole- and electron-like pockets must contribute. These general considerations give a useful intuition into the form of the Fermi surface found in our antiferromagnetic DFT calculations shown in Fig.~\ref{fig:fig1}(d), where the reconstructed Fermi surface of BaFe$_2$As$_2$  is found to consist of tiny 2D electron-like pockets centred exactly on high-symmetry axes, and other small 3D hole- and electron-like pockets.

Turning now to the experimental determination of the electronic structure using ARPES, the overview Fermi surface map in Fig.~\ref{fig:fig1}(e) obtained at 150~K, above $T_N$, reveals a structure typical of non-magnetic Fe-based superconductors, with three hole pockets at the zone centre and two electron pockets at the Brillouin zone edge \footnote{Based on a combination of measurements in different geometries, our understanding is that all three hole pockets cross the Fermi level at 150~K, consistent with non-magnetic DFT and Ref.~\cite{Pfau2019PRB}}. The measurements are strongly modulated by matrix elements effects \cite{Brouet2012}, but are qualitatively similar to the calculated Fermi surface, except that the sizes of the pockets are smaller than in the calculation \cite{Fedorov2018_arxiv,Borisenko2015NPhys}, and the features are broadened due to the elevated temperatures. However, at low temperatures, the overview Fermi surface on a twinned sample in Fig.~\ref{fig:fig1}(f) shows a completely different structure, prominently featuring four sharp bright spots around each high-symmetry point. These bright spots are in fact tiny electron-like Fermi surfaces, too small to be resolved in the measurement, and correspond to the tiny 2D tubes found in the DFT calculation. Notably, however, the calculation contains only two such tubes in the $\mathrm{\bar{\Gamma}-\bar{X}}$ direction only, whereas four bright spots are observed experimentally due to the superposition of intensity from the two domains. This illustrates the necessity of extracting ARPES data from mono-domain samples to fully probe the intrinsic anisotropies of the electronic structure. In Fig.~\ref{fig:fig1}(g,h) we show overview Fermi surface maps measured on a sample under an applied strain; the data are taken on the same sample and in the same measurement geometry, but with the sample rotated by 90$^\circ$ (see Methods section). Here it is confirmed that, as in the calculations, in one domain there are only two bright spots, i.e. two tiny tube-like electron pockets. Moreover, the total absence of any spots in the other orientation, despite their intrinsic brightness in this geometry, indicates that this sample must be close to being 100$\%$ detwinned. 

Much less brightly, one can also observe some faint intensity in replica bands, indicated by the circles in Fig.~\ref{fig:fig1}(g). This intensity cannot be ascribed to bulk electronic structure, but rather is a surface effect: after cleavage, the surface is terminated by Ba atoms, but not all atomic sites are occupied. This leads to possible surface reconstructions \cite{vanHeumen2011PRL}, of which a $\sqrt{2}\times{}\sqrt{2}$ reconstruction seems to best account for our observations. However, importantly this intensity is weak, and localised away from the high-symmetry points of interest. We are thus confident in ascribing the principle intensity observed to bulk electronic structure, though a contribution from surface-derived states even on the high-symmetry axes is hard to fully exclude \cite{vanHeumen2011PRL,Jensen2011PRB}. 

\begin{figure*}
	\centering
	\includegraphics[width=\linewidth]{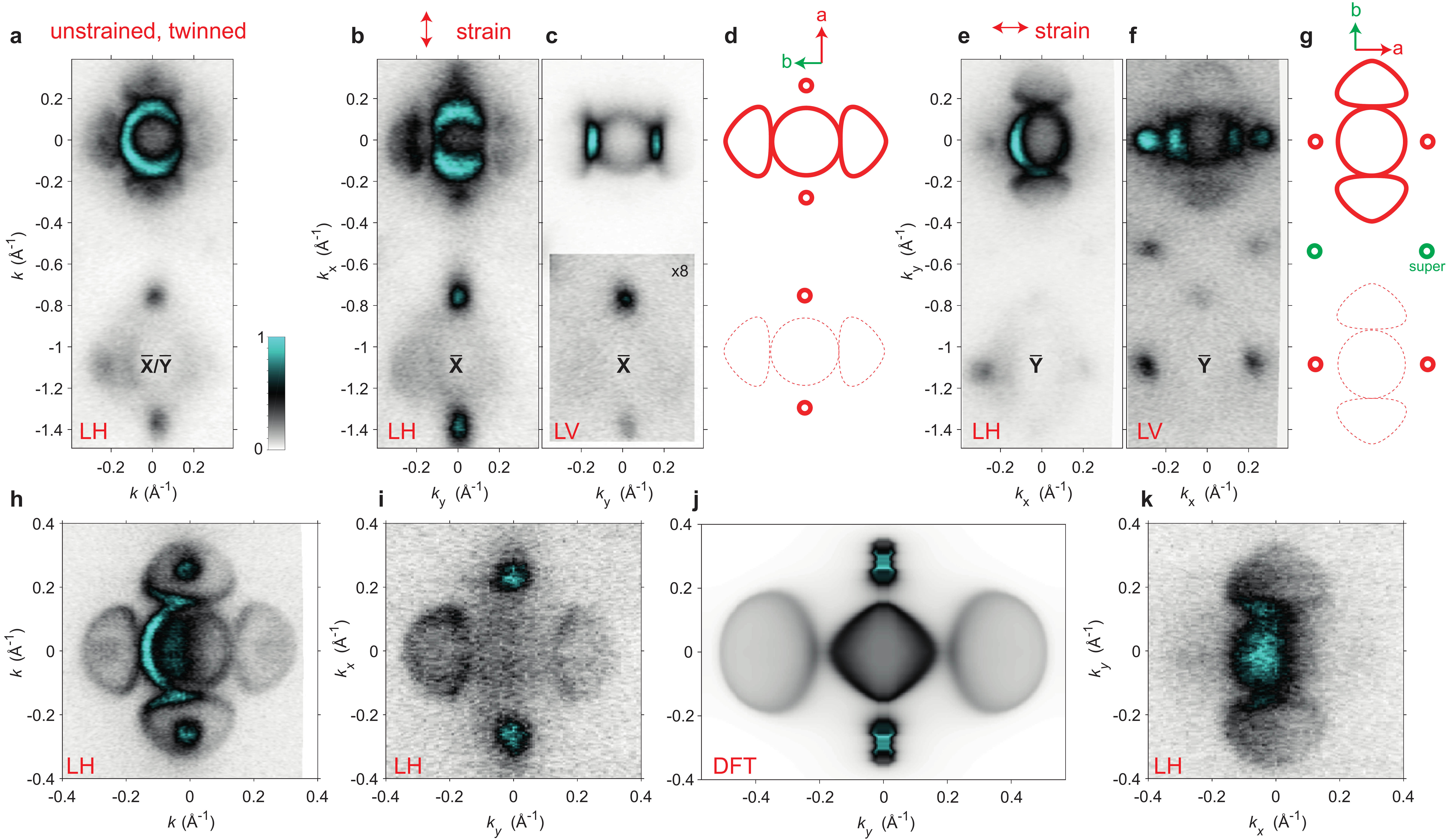}
	\caption{\textbf{Low-temperature Fermi surface maps before and after the application of strain.} (a) Twinned Fermi surface, measured using 62 eV photons in LH polarisation before applying strain to the sample. (b,c) Fermi surfaces measured after straining (detwinning) in $\mathrm{\bar{\Gamma}-\bar{X}}$ alignment in both LH and LV polarisations, and (d) schematic \textit{in-plane} Fermi surfaces. (e,f,g) Equivalent for $\mathrm{\bar{\Gamma}-\bar{Y}}$ alignment. h) Fermi surface map measured with 25 eV photons on a twinned sample. i,k) Equivalent measurement on the detwinned samples in the two orientations, and (j) the $k_z$-collapsed Fermi surface calculated by DFT in the stripe AFM phase for comparison. All data measured at 7-10~K.}
	\label{fig:fig2}
\end{figure*}

A geometric detail of BaFe$_2$As$_2$ is that the body-centred symmetry of the unit cell imposes a screw symmetry along $k_z$ at the corner of the Brillouin zone (Fig.~\ref{fig:fig1}(a,b)), similar to Sr$_2$RuO$_4$ \cite{Bergemann2003reviewSr2RuO4}, but notably different to the simpler primitive Brillouin zones of LiFeAs and FeSe. This adds complexity to the formal labelling of the Brillouin zone, which becomes further complicated by the stacking of the antiferromagnetic Brillouin zones (Fig.~\ref{fig:fig1}(c)). Since in experiments we always find a degree of $k_z$ integration, in this paper we follow the convention of Yi \textit{et al.} \cite{Yi2011PNAS} to index our data in terms of a simplified effective \textit{surface} Brillouin zone, using $\mathrm{\bar{X},\bar{Y}}$ to denote the corners of the non-magnetic Brillouin zone independent of $k_z$, and using $\mathrm{\bar{X}}$ and $\mathrm{\bar{Y}}$ separately for the $a$ and $b$ axes ($a>b$) in the case of data on detwinned samples.

\subsection{Detailed Fermi surface structure in one domain}

With this broad overview of the Fermi surface reconstruction in mind, we now turn to the high-resolution measurements in Fig.~\ref{fig:fig2}. In Fig.~\ref{fig:fig2}(a), the sample was prepared on the detwinning rig, but the measurement was taken before any strain was deliberately applied \footnote{Some residual strain due to differential thermal expansion may have been present, but it seems to have been insufficient to significantly detwin the sample in this case.}. The data in Fig.~\ref{fig:fig2}(a) clearly contains contributions from both domain orientations, i.e. the sample was twinned. We then tuned the strain to induce a transfiguration of the sample, from being fully twinned to being almost fully detwinned, entirely \textit{in situ}. These detwinned measurements in Fig.~\ref{fig:fig2}(b-g), taken in two orthogonal linear polarisations for completeness, allow the unmasking of the finer details of the Fermi surface in the AFM ground state. 

According to the DFT calculation, the larger pockets are both predicted to be 3D, and centred at different $k_z$. At the chosen photon energy of 62 eV, the inner, almost circular hole band dominates the intensity in Fig.~\ref{fig:fig2}(a). Since the photon energy is linked to the $k_z$ of the states probed in photoemission measurements, this indicates that the effective $k_z$ here is tuned to an intermediate point bisecting the hole-like Fermi surface in Fig.~\ref{fig:fig1}(d) (see Ref.~\cite{Fedorov2018_arxiv} for detailed discussion of $k_z$-dependence in BaFe$_2$As$_2$). However, the fact that the outer taco-shaped electron-like band can be simultaneously observed in the measurements is already testament to the fact that there is a substantial degree of $k_z$-uncertainty in photoemission, due to both the finite escape depth of the photoelectron, and because the final state dispersions are in general unknown. Moreover, due to the variation of $k_z$ with in-plane momentum, the measurements at $\mathrm{\bar{X}}$ and $\mathrm{\bar{Y}}$ probe $k_z$ close to the $k_z$=0 plane, and so only show the bright spots. Still, this geometry is convenient for relatively precise determinations of the \textit{in-plane} shape of the Fermi surfaces, which are represented in the schematics in Fig.~\ref{fig:fig2}(d,g). 

The outlines of all the Fermi surfaces, especially the taco-shaped pockets, are seen most clearly in Fig.~\ref{fig:fig2}(h,i,k) where we present measurements at 25 eV photon energy. In Fig.~\ref{fig:fig2}(h) in particular, the highest resolution measurements in this paper, one can begin to see that the ``bright spots" have an internal structure, and are in fact tiny electron-like pockets. The asymmetry of the spectral weight with respect to $k_x$ in this geometry closely resembles the one-step photoemission calculations in Ref.~\cite{Derondeau2016PRB} and reflects a complex multiorbital character of the reconstructed Fermi surfaces. The $k_z$ averaging effect results in the observation of mainly the outlines of extremal areas of the pockets, but also a continuum of spectral weight inside.  

In order to make a more quantitative comparison to DFT calculations, in Fig.~\ref{fig:fig2}(j) we plot the Fermi surface obtained by collapsing the AFM DFT calculations over all $k_z$ values. While the observed Fermi surfaces in Fig.~\ref{fig:fig2}(i,k) are qualitatively similar to the calculation, the overall size of the Fermi pockets are significantly smaller than the DFT calculation shown in Fig.~\ref{fig:fig2}(j). It is well known that the Fermi surfaces of near-stoichiometric Fe-based superconductors in their tetragonal phases typically show smaller hole- and electron- pockets than predicted by DFT \cite{Ortenzi2009PRL,Carrington2011Review,Borisenko2015NPhys,Watson2018PRB_NaFeAs,Fedorov2018_arxiv}, as we also found in Fig.~\ref{fig:fig1}(a) above $T_N$, but here we see that similar logic applies also to the reconstructed Fermi surface in the AFM ground state.

\subsection{Anisotropic band dispersions in the AFM phase}

 \begin{figure*}[t]
 	\centering
 	\includegraphics[width=0.7\linewidth]{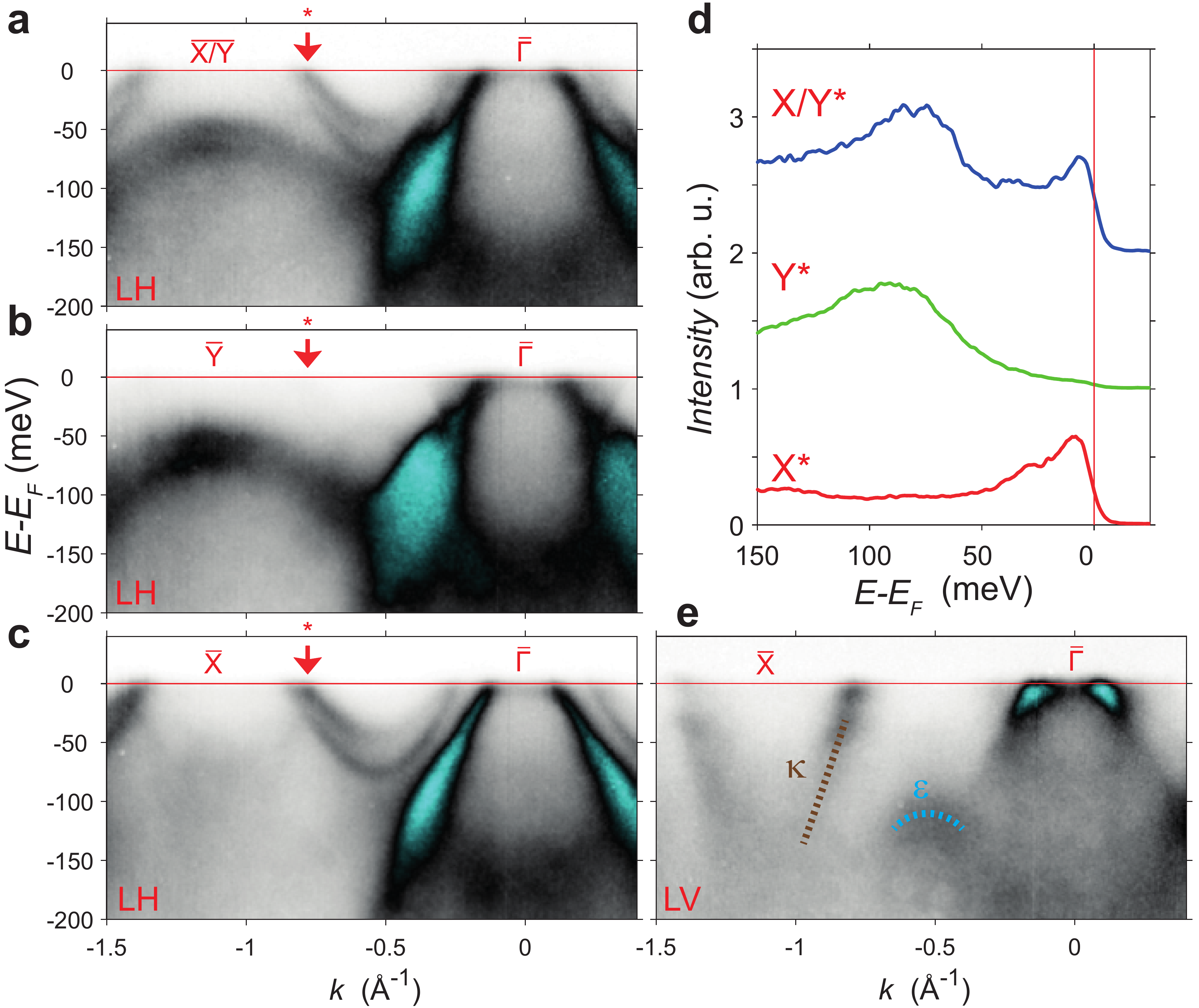}
 	\caption{\textbf{Distinct band dispersions along} $\mathbf{\mathrm{\bar{\Gamma}-\bar{X}}}$ \textbf{and} $\mathbf{\mathrm{\bar{\Gamma}-\bar{Y}}}$\textbf{.} (a) High-symmetry dispersion before applying strain and (b,c) equivalent measurements on the same sample after straining, which is now in a detwinned state, with the sample oriented along $\mathrm{\bar{\Gamma}-\bar{Y}}$ and $\mathrm{\bar{\Gamma}-\bar{X}}$ respectively. The difference between the two orientations is particularly prominent at the $k$-point labelled by the starred arrow, from which the EDCs in (d) are extracted, showing an apparent $\sim$70 meV energy scale. These measurements are qualitatively similar to those reported by Yi \textit{et al.}, but see Fig.~\ref{fig:fig4} and main text for revised interpretation. (e) Measurement with equivalent geometry to (c) but with LV light polarisation, which highlights other states including a band ($\kappa$) with mainly $d_{xy}$ character. The states labelled $\kappa$ and $\epsilon$ are also marked in Fig.~\ref{fig:fig4}(c) and referred to in the main text. All measurements performed using $h\nu$ = 62~eV at 7-10~K.}  
 	\label{fig:fig3}
 \end{figure*}

 Having established the shape of the Fermi surface, we now turn to the electronic dispersions along the high-symmetry directions. The measurements on the twinned sample before applying strain in Fig.~\ref{fig:fig3}(a) are difficult to interpret, since the $\mathrm{\bar{\Gamma}-\bar{X}}$ and $\mathrm{\bar{\Gamma}-\bar{Y}}$ dispersions are observed simultaneously. After the strain is applied, however, the measurements on the now detwinned sample in Fig.~\ref{fig:fig3}(b,c) reveal the very distinct band dispersions along the $\mathrm{\bar{\Gamma}-\bar{X}}$ and $\mathrm{\bar{\Gamma}-\bar{Y}}$ directions at low temperatures \cite{Yi2011PNAS,Kim2011PRB_Kwon}. Near $\mathrm{\bar{Y}}$ in Fig.~\ref{fig:fig3}(b), a relatively flat band is observed which remains fully occupied, whereas near $\mathrm{\bar{X}}$ a band crosses the Fermi level in Fig.~\ref{fig:fig3}(c), creating the tiny 2D tube-like Fermi surface pocket in combination with the an electron-like dispersion with $d_{xy}$ character. This latter band is seen only in the opposite polarisation, labelled $\kappa$ in Fig.~\ref{fig:fig3}(e)) \cite{Jensen2011PRB} \footnote{The lack of hybridisation between the two bands observed in opposite polarisations along $\mathrm{\bar{\Gamma}-\bar{X}}$ is the origin of the `Dirac' character of the tiny electron-like Fermi surfaces.}.  Focusing on a particular $k_{||}$ marked by the red arrows, we extract the Energy Distribution Curves (EDCs) shown in Fig.~\ref{fig:fig3}(d). This analysis reveals a large ~70 meV energy scale between the primary band dispersions in the two directions. It was previously proposed by Yi \textit{et al.}, that this large energy scale reflects an orbital ordering underlying the magnetic phase in the pure BaFe$_2$As$_2$, which was also observable independent of magnetic order in a Ba(Fe$_{0.975}$Co$_{0.025}$)$_2$As$_2$ sample where $T_s$ and $T_N$ were split \cite{Yi2011PNAS}. Very recently, Pfau \textit{et al.} took this view a step further by trying to account for the low temperature dispersions by applying such an orbital ordering, and subsequently backfolding (but not hybridising) the dispersions \cite{Pfau2019PRB}. However, here we take a contrasting view, by placing the AFM ordering in the driving seat. 

 \begin{figure*}
	\centering
	\includegraphics[width=\linewidth]{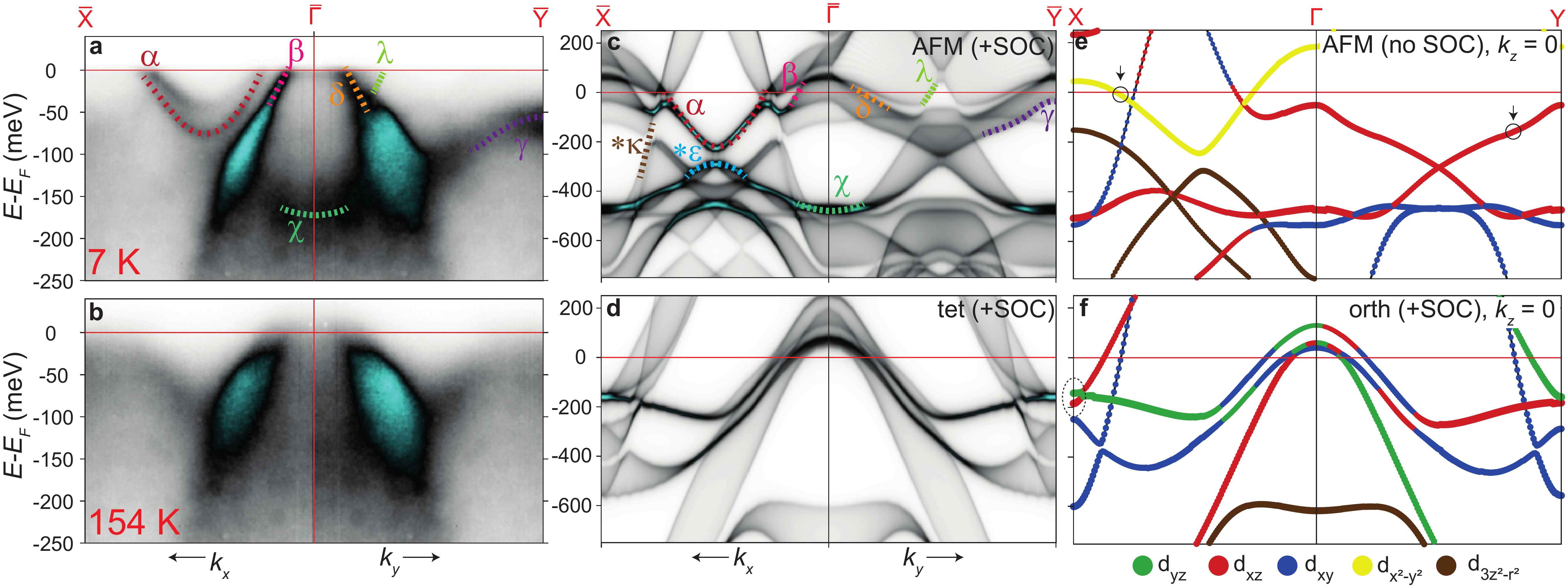}
	\caption{\textbf{Comparisons of data above and below $\mathbf{T_N}$ with DFT.} (a) Stitched ARPES measurements along the orthogonal high-symmetry axes (same data as Fig.~\ref{fig:fig3}b,c) and (b) equivalent measurements above $T_N$. (c) DFT calculation in the AFM ground state, with band dispersions averaged along the $k_z$ axis to generate a spectral function. Note that in the experiments, the intensities are very different, since not all eigenvalues will have equal spectral weight in the magnetic phase, and also there are severe matrix element effects. Nevertheless, good correspondence can be seen for certain states, as labelled. Note that the starred states, labelled $\epsilon$ and $\kappa$, are found in LV polarisation in Fig.~\ref{fig:fig3}(e). (d) Equivalent projection of DFT calculations in the tetragonal phase. (e) Calculated dispersion at $k_z$=0 only in the AFM phase, with orbital characters shown. Small circles note the bands which mainly contribute to the EDCs shown in Fig.~\ref{fig:fig3}(d). (f) Calculated dispersions at $k_z$=0 in the orthorhombic phase. The orthorhombicity lowers the $d_{xz}/d_{yz}$ degeneracy, especially noticeable at $X$ as highlighted by the dashed ellipse.}
	\label{fig:fig4}
\end{figure*}

From the DFT perspective, the stripe AFM of BaFe$_2$As$_2$ is not a weak-coupling SDW; that is to say, the low temperature band dispersions cannot simply be accounted for by backfolding bands from the normal phase and weakly hybridising them. Rather, the magnetic order completely reconstructs the electronic structure, modifying the dispersions, orbital characters, and even the number of bands at low energies. For example, in the DFT calculation in Fig.~\ref{fig:fig4}(f) in the orthorhombic phase, there are three hole pockets at the $\Gamma$ point. After backfolding the two electron dispersions, one might thus expect to find five states at low energies at the $\Gamma$ point of the AFM phase; however there is only one band found within 100 meV of the Fermi level at the $\Gamma$ point in the AFM DFT calculation in Fig.~\ref{fig:fig4}(e). The AFM thus completely reconstructs the band dispersions, with only a select few band dispersions surviving in a recognisable form (such as the $d_{xy}$ dispersion at X in Fig.~\ref{fig:fig4}(e)). Moreover, as the $\mathrm{\Gamma}$-$\mathrm{X}$ and $\mathrm{\Gamma}$-$\mathrm{Y}$ directions correspond to AFM and FM spin alignments respectively, the band dispersions along $\mathrm{\Gamma}$-$\mathrm{X}$ and $\mathrm{\Gamma}$-$\mathrm{Y}$ will substantially differ in the AFM phase. 

At first glance, the experimental dispersions in Fig.~\ref{fig:fig4}(a,b) do not appear to be so drastically reconstructed in the low temperature AFM phase, but in fact there are several signatures that point to a strong effect of the AFM order on the electronic structure. For instance, the states labelled $\chi$ in Fig.~\ref{fig:fig4}(a,c), and $\epsilon$ in Fig.~\ref{fig:fig3}(e) are new states found only in the AFM phase, with no counterpart in the non-magnetic phase, but can be identified in the AFM calculation. The band labelled $\alpha$ in Fig.~\ref{fig:fig4}(a) was previously attributed to a shifted $d_{yz}$ dispersion \cite{Yi2011PNAS,Pfau2019PRB}. However, its appearance as an almost-symmetric V-shape along $\mathrm{\Gamma}$-$\mathrm{X}$ cannot be reconciled with the much flatter dispersion of the $d_{yz}$ band in Fig.~\ref{fig:fig4}(d,f). Rather, a matching V-shaped dispersion is found in the AFM DFT calculations in Fig.~\ref{fig:fig4}(c,e), which is qualitatively distinct from any dispersion in the non-magnetic phase. We already showed that the Fermi surface is in qualitative agreement with the DFT prediction, but it is worth emphasising that, if one were to simply backfold the non-magnetic bands, one would inevitably find several more small pockets than are actually found. The comparison between the DFT and experiment requires some careful thought, as some band shifts are required. Moreover, not all the eigenvalues in the AFM calculation will give significant spectral weight in ARPES, and the $k_z$ integration is always relevant to the observations. However, overall we can make enough one to one correspondences in Fig.~\ref{fig:fig4}(a,c) that we can be satisfied that, broadly speaking, the DFT perspective is correct: the AFM ordering strongly reconstructs the electronic structure, across the whole bandwidth, and dominates over any orbital or nematic ordering.

\begin{figure*}[t]
	\centering
	\includegraphics[width=\linewidth]{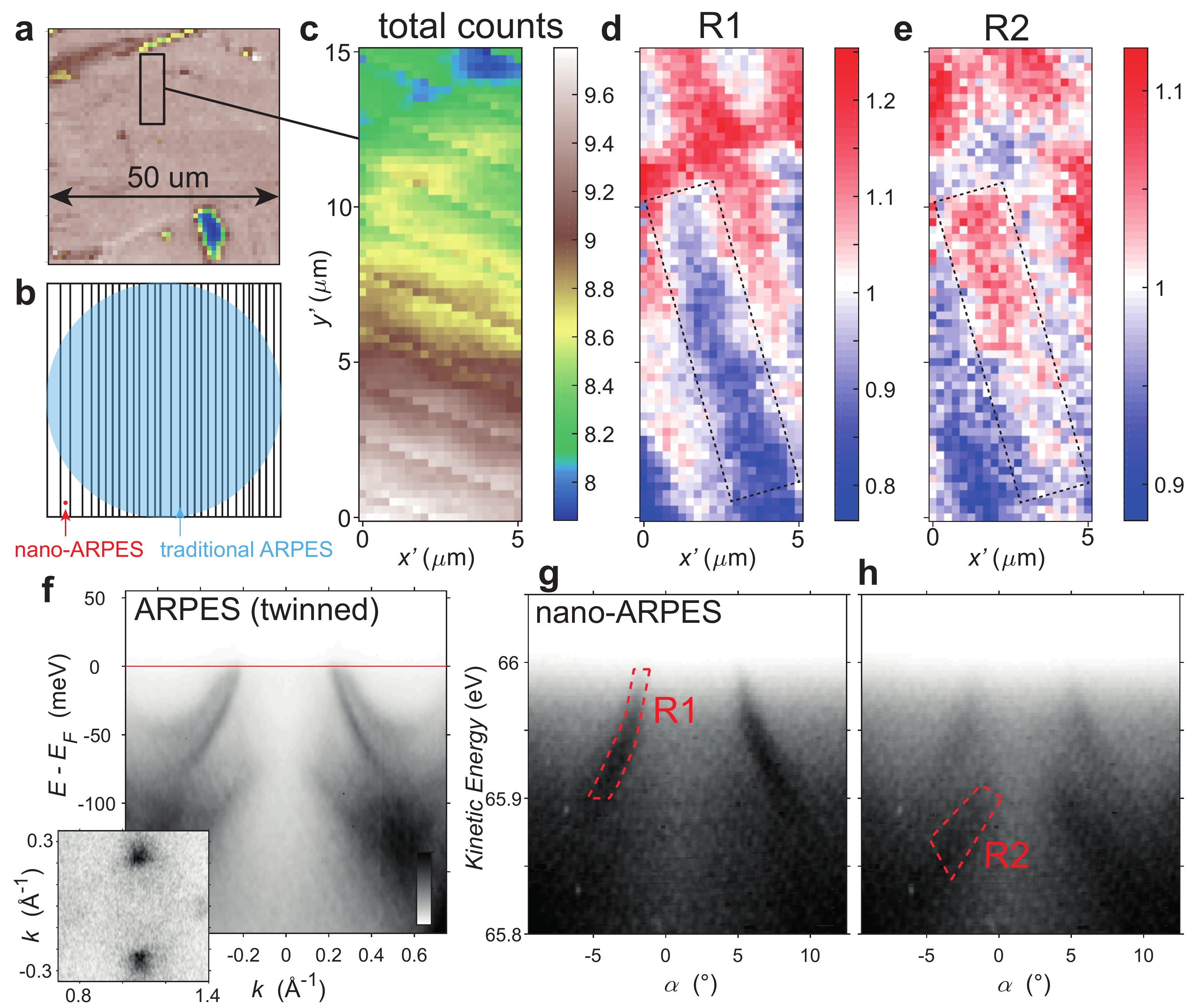}
	\caption{\textbf{Mapping of real space and reciprocal space structures with nano-ARPES.} (a) Total intensity map over a 50$\mu$m region, from which a 5$\times$15$\mu$m area was selected for detailed study. (b) Illustration of the relative beam size of nano-ARPES versus conventional ARPES. Vertical lines represent typical domain sizes in BaFe$_2$As$_2$. c) Total intensity map of the detailed area, which is largely featureless (the stripes in this image are artefacts caused by the periodic ``top-ups" of the synchrotron, which weakly modulates the incident flux). Note that the $x',y'$ spatial axes here are rotated with respect to the crystallographic axes. The hidden domain structures are revealed by analysing the intensity within (d) region of interest R1 and (e) region of interest R2 (normalised by the total count at each pixel). Dotted black line is a guide to see the reversal of the domain pattern. (f) Reference measurements obtained with traditional ARPES on a twinned sample, inset showing the Fermi surface. (g,h) Effective single-domain spectra, extracted from a filtering technique, summing intensity at the pixels with the top (g) and bottom (h) 10$\%$ contrast of region R1. Since the spatial mapping data is analysed on the raw, not $k$-scaled data, we plot the raw data here. Note that the regions of interest are defined including a symmetric area at positive angles, which is not shown. All nano-ARPES data measured at 25~K.}
	\label{fig:nanofigure}
\end{figure*}

Returning to the $\sim$ 70 meV energy scale as originally found by Yi \textit{et al.} \cite{Yi2011PNAS}, the corresponding bands are identified by circles in the DFT calculation in Fig.~\ref{fig:fig4}(e), which suggests that the lower band, found near $\mathrm{\bar{Y}}$, is likely to have $d_{xz}$ orbital character, but the upper dispersion $\mathrm{\bar{X}}$ is predicted to have mainly $d_{x^2-y^2}$ character and only minority $d_{yz}$ character.  Thus we argue that it is not possible to cleanly ascribe this energy scale as being due to ``orbital ordering" lifting the degeneracy of $d_{xz}$and $d_{yz}$ states, firstly because the orbital content changes (at least, in this DFT calculation), secondly because it is only applicable to a particular $k_{||}$ and our data is not generally suggestive of a persistent orbital polarisations at other points, and thirdly because all of the changes in the bandstructure are inseparable from the dominant effect of the magnetic ordering.  Furthermore, our temperature-dependent data (shown in supplementary information, SI) demonstrates that the majority of the band structure changes occur relatively abruptly at $T_s=T_N=$~137~K. This is in line with the observation of a first-order phase transition in thermodynamic measurements \cite{Boehmer2015NComms}. However, since the transition is first order, the procedure of tracking bands from the non-magnetic phase into the ordered phase \cite{Yi2011PNAS,Pfau2019PRB} is questionable in BaFe$_2$As$_2$, especially given the high scattering rate at temperatures around $T_N$.

Of course, there is a significant discrepancy between the energy scales of the data and the DFT calculations. The experiments show an overall renormalisation of the bandwidth of a factor of $\sim$3 \cite{Kim2011PRB_Kwon}, in both the non-magnetic and ordered phases, which is usually attributed to strong on-site electron correlations. Dynamical Mean Field Theory (DMFT) has been a widely used tool to treat the strong local Hund's and Coulomb terms within the Fe $3d$ orbitals, overcoming the limitations of DFT in describing such interactions and achieving much better correspondence with experiments on the overall energy scales of the dispersions \cite{Watson2017DMFT,EvtushinskyNaFeAsDMFT}. While relatively few studies have treated the AFM phase within DMFT or similar techniques, our data appear to be well matched in terms of overall energy scales and anisotropies with the magnetic DMFT calculation of Yin \textit{et al.} \cite{Yin2011NPhys}. However, we find that other than the overall bandwidth renormalisation, DFT in the stripe AFM phase offers a good description of the band dispersions, as is evidenced by the agreement between calculated and measured Fermi surfaces in Fig.~\ref{fig:fig3}.

\subsection{nano-ARPES: band dispersions from within individual domains}

The measurements above have re-emphasized the value of performing measurements on ``detwinned" samples, but the application of external strain on the samples is not ideal, for both experimental reasons (i.e. the possible bending and cracking of the sample surface, and strain inhomogeneity) and also more conceptual reasons (for example, because of possible changes induced by the strain due to the magnetoelastic coupling). Ideally, therefore, one would like to be able to measure ARPES from within a single domain of an unstrained sample. In the following we demonstrate that this is now feasible using nano-ARPES, verifying the ``detwinned" results obtained on strained samples and opening the door to many exciting experimental opportunities.   

Previous studies of domain formation in BaFe$_2$As$_2$ have shown that in unstrained samples, domains typically form as long stripes, with straight boundaries separated by a characteristic spacing of a few microns \cite{Tanatar2010PRB,Fisher2011review,Kirtley2010PRB}. This inevitably means that a superposition of intensity from two domains is detected in a traditional ARPES set-up, with a typical beam spot of 50 $\mu$m or larger. Developments in laser-ARPES have reduced the spot size to a few microns \cite{Iwasawa2017Ultramicroscopy}, and in a recent study Schwier \textit{et al.} were able to visualise some spatial variation of the intensities of hole pocket dispersions near the $\Gamma$ point in FeSe, which was associated with domain structures \cite{Schwier2019}. However, the capability to combine excellent spatial resolution and full $k$-space mapping has become recently accessible to state-of-the-art nano-ARPES, where the beam spot can be focused to better than 700 nm in diameter. The greatly decreased spot size in our nano-ARPES apparatus comes at the price of a substantially lower count rate, necessitating measurements with lower energy resolution, and also a lower signal/noise ratio. Nevertheless, it provides an excellent opportunity to study electronic anisotropies in a strain-free measurement, resolving any questions over the influence of the external strain in the traditional ARPES measurements. 

Our chosen measurement geometry probes a dispersion through the $\mathrm{\bar{X},\bar{Y}}$ points at 70 eV in LH polarisation. Due to the selection rules in this geometry, each domain mainly contributes just one band: as can be seen in our traditional ARPES reference data in Fig.~\ref{fig:nanofigure}(f),  the band from one domain approaches the Fermi level to create the tiny spot-like pockets; in the other domain, a band is observed at higher binding energies (with correspondingly higher linewidth due to electron-electron scattering), reaching a maximum approximately 100 meV below $E_F$ at the high-symmetry point, where the intensity also vanishes. 

Since these two bands are well-separated in both energy and momentum, they function as convenient indicators of the contribution of each domain, at any given spot on the sample. They are roughly equivalent in total intensity, so the spatial map of the integrated intensity in Fig.~\ref{fig:nanofigure}(c) is largely uniform, showing no clear domain-like structures. However, the underlying domain structure is revealed in Fig.~\ref{fig:nanofigure}(d), where the colour scale is proportional to the intensity of a region of interest corresponding to the upper dispersion (R1 in Fig.~\ref{fig:nanofigure}(g)). This reveals straight, parallel domains, with a lateral spacing of $\sim$2~$\mu$m, consistent with the structures observed in the literature \cite{Tanatar2010PRB,Fisher2011review,Kirtley2010PRB}. The upper part of the spatial map includes a probable crack or defect in the sample, also visible in the total counts in Fig.~\ref{fig:nanofigure}(c). This feature seems to locally pin a particular domain, presumably due to some residual stress. However, the main domains appear to run straight through this. We may speculate that a crack or terrace-like feature may be present only at the surface, while the main domains presumably extend vertically deep into the sample. An important test is that the domain contrast should flip when analysing the spatial map for the lower R2 regions of interest. This is confirmed in Fig.~\ref{fig:nanofigure}(e), where the intensity pattern is reversed compared with Fig.~\ref{fig:nanofigure}(d). 

This analysis allows us to precisely position the beam spot inside either domain, allowing for the unprecedented collection of band dispersions from within a single domain (see SI for spectra obtained at a fixed light spot). Alternatively, one can build up a mono-domain spectrum by summing intensity from pixels in the spatial map which have the highest contrast for the desired domain; the results of this filtering technique are shown in Fig.~\ref{fig:nanofigure}(g,h), which indeed reveals ``detwinned-like" dispersions. This filtering technique generates effective mono-domain dispersions, while overcoming the experimental challenges of sample drift and degradation. 

From a technical perspective, these nano-ARPES results push the boundaries of achieving detailed band structure information from within micron-size domains. Our results are very promising for further nano-ARPES studies of other quantum materials which exhibit magnetic or structural domains in their ground states, allowing the spatial mapping of electronic anisotropies. The combination of spatial and momentum resolutions demonstrated in this paper is also beginning to find more widespread applications to systems where $\sim$ micron length scales are relevant, such as in the study of crystals where cleavage yields multiple surface terminations \cite{Noguchi2019Nature,Mazzola2018PNAS} or only small regions with flat surfaces \cite{Schroeter2018_arxiv,Valbuena2019PRB}, systems exhibiting phase separation, and patterned devices. We hope that the data presented above can therefore serve as a proof-of-concept study on a material which is well known to the quantum materials community, to motivate further use of this technique. 

Beyond the technical achievement, however, there are two important scientific points. Firstly, our nano-ARPES results are consistent with the traditional ARPES measurements on detwinned samples: we have shown that the bands forming the bright spots in the Fermi surface derive from different domains, with each domain giving just two bright spots (tiny electron pockets). Thus, we can lay to rest any worry that the ARPES measurement on detwinned samples are unduly influenced by the application of the external strain, and confirm that the ARPES data here and in the literature on detwinned samples can indeed be safely interpreted as the intrinsic spectral function of one domain (for a fully detwinned sample). Secondly, our measurements act as a reminder that, within the 50 $\mu{}$m beam spot on a nominally clean single crystal, there can be several cracks, inclusions, inhomogeneities, as well as domain structures, which are all integrated into the traditional ARPES measurement. Such microscopic information, on a length scale which spans the regimes of STM and traditional ARPES, is important for interpretation of the spectroscopy data obtained with a macroscopic beam spot.

\section{Discussion}

The Fermi surface we have measured closely resembles that deduced by quantum oscillations \cite{Terashima2013PRL}. Moreover, the small pockets of compensated carriers, including a small number of highly-mobile carriers from the quasi-2D tube-like electron pockets with Dirac-like dispersions,  would also provide a consistent explanation of the magnetotransport data \cite{Ishida2011PRB}. While the DFT prediction differs from this in terms of the size of the pockets, the topology of the band is found to be consistent. Our study also resolves several inconsistencies in the previous ARPES data: Yi \textit{et al.} \cite{Yi2011PNAS} indicated several more Fermi surfaces than were detected by quantum oscillations or predicted by DFT, while other previous studies also lacked the resolution to be precise on the nature of the Fermi surface \cite{Shimojima2010PRL,Richard2010PRL,Kim2011PRB_Kwon,Jensen2011PRB}. Only very recently, Fedorov \textit{et al.} \cite{Fedorov2018_arxiv} reported Fermi surfaces measured on twinned samples that, for the first time, appeared broadly consistent with DFT calculations on BaFe$_2$As$_2$. Here, with high-resolution measurements on detwinned samples, we cement the experimental Fermi surface of BaFe$_2$As$_2$ in one domain, and reveal how the magnetic ordering anisotropically reconstructs the electronic structure over the whole bandwidth.

In another very recent study, Pfau \textit{et al.} described the ground state dispersions of BaFe$_2$As$_2$ by applying band shifts which they associated with nematic order, and then backfolded the bands, i.e. giving prominence to the orbital ordering due to its apparently large energy scale and taking a weak coupling approach to the AFM \cite{Pfau2019PRB}. Our \textit{ab-initio}-based approach gives us the opposite philosophy: we argue that the AFM severely reconstructs the electronic structure over the whole bandwidth, and magnetism is firmly in the driving seat. The low temperature Fermi surface provides strong support for our approach, since it is nearly correct in the AFM DFT calculation, but any simple backfolding procedure would lead to far more pockets than are actually observed \cite{Pfau2019PRB}. However, we note that the scenario is somewhat different in NaFeAs, where due to the much lower $T_N$ and much weaker ordered moment (0.09 $\mu_B$ rather than 0.87 $\mu_B$ in BaFe$_2$As$_2$ \cite{Dai2015review}), the magnetic ordering is more SDW-like and the approach of understanding the ground state Fermi surface in terms of backfolded bands is more fruitful \cite{Watson2018PRB_NaFeAs}. The presence of two tiny tube-like Fermi pockets, appearing as two bright spots along the $\mathrm{\bar{\Gamma}-\bar{X}}$ direction in the measurements, gives a strongly unidirectional character to the measured Fermi surface of BaFe$_2$As$_2$, and is reminiscent of the as-yet unexplained observation of only the electron pocket oriented along the $a$ axis in detwinned measurements of the nematic phase of FeSe \cite{Watson2017NJP}. We speculate that there is indeed a connection between the two observations, but we leave this to future work to unravel. 

In conclusion, we have used high-resolution ARPES measurements of detwinned BaFe$_2$As$_2$ to revisit the ground electronic state in the magnetically ordered phase. We have shown that the Fermi surface in the AFM phase includes both 3D pockets and tiny quasi-2D tubes, closely matching the prediction of DFT in the AFM phase, though with smaller in-plane size. The observed low temperature dispersions cannot be understand by simply shifting and backfolding the high-temperature dispersions, but rather the magnetic order reconstructs the electronic structure over the whole bandwidth. Our measurements of samples detwinned by the application of a mechanical strain were corroborated by nano-ARPES measurements, in which the spot scanned over individual domains and single-domain spectra were obtained on an unstrained sample. Overall, the revision of the ARPES evidence presented here puts the spotlight back onto magnetic interactions as the main ingredient in the phase diagrams based on BaFe$_2$As$_2$, the archetypal parent compound of Fe-based superconductors.

\section{Acknowledgements}

%\begin{acknowledgments}
We thank S.~Backes, S.~V.~Borisenko, N. B. M. Schr\"{o}ter, M.~Eschrig, R.~Valenti and C.~Cacho for useful discussions. We thank Diamond Light Source for access to beamline I05 (proposal numbers SI15074,SI19041) that contributed to the results presented here. The work at IFW was supported by the Deutsche Forschungsgemeinschaft (DFG) through the Priority Programme SPP1458. L. C. R. is supported by an iCASE studentship of the UK Engineering and Physical Sciences Research Council (EPSRC) and Diamond Light Source Ltd CASE award. SA thanks the DFG for funding (AS 523\textbackslash{}4-1 \& 523\textbackslash{}3-1).

\subsection{Methods}
High quality single crystals of BaFe$_2$As$_2$ were grown by the self flux technique \cite{Aswartham2011}. For the detwinned measurements, a sample with approximate dimensions of 1500 $\times$ 1200 $\times$ 50 $\mu$m  with uniform thickness and regular facets was selected and mounted across the plates of the horseshoe-shaped device, aligned by eye (within $\sim 2^\circ$) such that the Fe-Fe direction was parallel with the direction of strain. Silver epoxy (Epo-Tek H27D) was used to mount the sample and also acted as a medium to transmit the strain into the sample. Due to the finite Poisson's ratio, the actual strain on the sample ought to be described by a full strain tensor, but for simplicity we refer to a unidirectional tensile strain. The photoelectron energy and angular distributions were analysed with a SCIENTA R4000 hemispherical analyser The angular resolution was 0.2$^\circ$, and the overall energy resolution was better than 10 meV. Nano-ARPES measurements were performed on a similar sample using 70 eV photon energy and SCIENTA DA30 analyser, at a temperature of 30~K and with a typical energy resolution of $\approx$30 meV chosen due to the much lower count rate, and angular resolution of $\sim$0.2$^\circ$. It was found that prolonged exposure ($>$30 minutes) of the sample to the focused beam spot at a fixed position caused local sample degradation, but in scanning mode (typically 1 minute/pixel) this was not a serious problem. Both ARPES and nano-ARPES measurements were performed at the I05 beamline at the Diamond Light Source, UK \cite{I05beamlinepaper}. The density functional theory (DFT) calculations were performed using the Wien2k code, as detailed in the SI. 

%\end{acknowledgments}

\section{Author contributions}

ARPES experiments were performed by MDW, LCR, HI, and TKK. Nano-ARPES experiments were performed by MDW, PD, LCR, DE, HI, and TKK. The acquisition of both data sets was enabled by innovations at the I05 beamline developed by PD, HI, MDW, TKK, and MH. Samples were grown by SA, SW and BB. MDW analysed the data, performed the DFT calculations and wrote the manuscript with TKK, with input from all co-authors. The overall project was led by MDW and TKK.    

\clearpage

%\bibliographystyle{Science}
%\bibliography{../BaFe2As2_bibliography}
%% PASTED BBL

\clearpage
\onecolumngrid
	
	\section{Supplementary Information} 
	
	\section{Density Functional Theory calculations}
	Density functional theory calculations were performed in the Wien2k code, using the generalised gradient approximation (PBEsol) exchange-correlation potential. A large number of $k$-points ($\sim$50000 points in the Brillouin zone) was used, in order to accurately plot the 3D Fermi surfaces in Fig 1. Note that in the magnetic case, the tiny tube-like pockets are very sensitive to the exact position of the chemical potential, and a different topology (e.g. closed 3D pockets) can be obtained by changing some parameters of the calculation (e.g. exact lattice constants, choice of exchange-correlation potential etc). The ordered moment in the magnetic phase was 1.93 $\mu_B$/Fe, an overestimate of the experimental case of 0.87 $\mu_B$/Fe, but this is a well-known problem in DFT modelling of the parent phases of Fe-based superconductors. Due to a technical issue, the projection of orbital characters in the AFM phase in Fig 4(e) of the main text was performed without including spin-orbit coupling; otherwise spin-orbit coupling is included on all sites. In the tetragonal phase, we used the lattice parameters $a$=3.9622, $c$=13.001, $z_{As}$= 0.10393 in space group $I4/nmm$ (139). In the AFM phase, we use the lattice parameters $a$=5.6157, $b$=5.5718, $c$=12.94240,  $z_{As}$=0.10375 in space group $Cccm$ (66). Note that Wien2k uses a rotated structural setting in this case, but here we use the conventional setting where the $c$ axis is out of the plane. The non-magnetic orthorhombic calculation was performed using the same lattice constants, but in the absence of magnetic ordering the space group is $Fmmm$ (69).    
	
	We note that in the recent paper of Fedorov \textit{et al.} \cite{Fedorov2018_arxiv}, some rigid band shifts were applied to obtain better agreement with the data. However, here we choose not to go for such tweaks, rather we only show the ``vanilla" DFT calculation, which already shows reasonable agreement. Our data provides a good target for future \textit{ab-initio} studies of the AFM phase, with dynamical correlation effects being a key component. 
	
	\clearpage	
	
	\section{Temperature-dependent ARPES measurements}
	\begin{figure*}[h]
		\centering
		\includegraphics[width=0.97\linewidth]{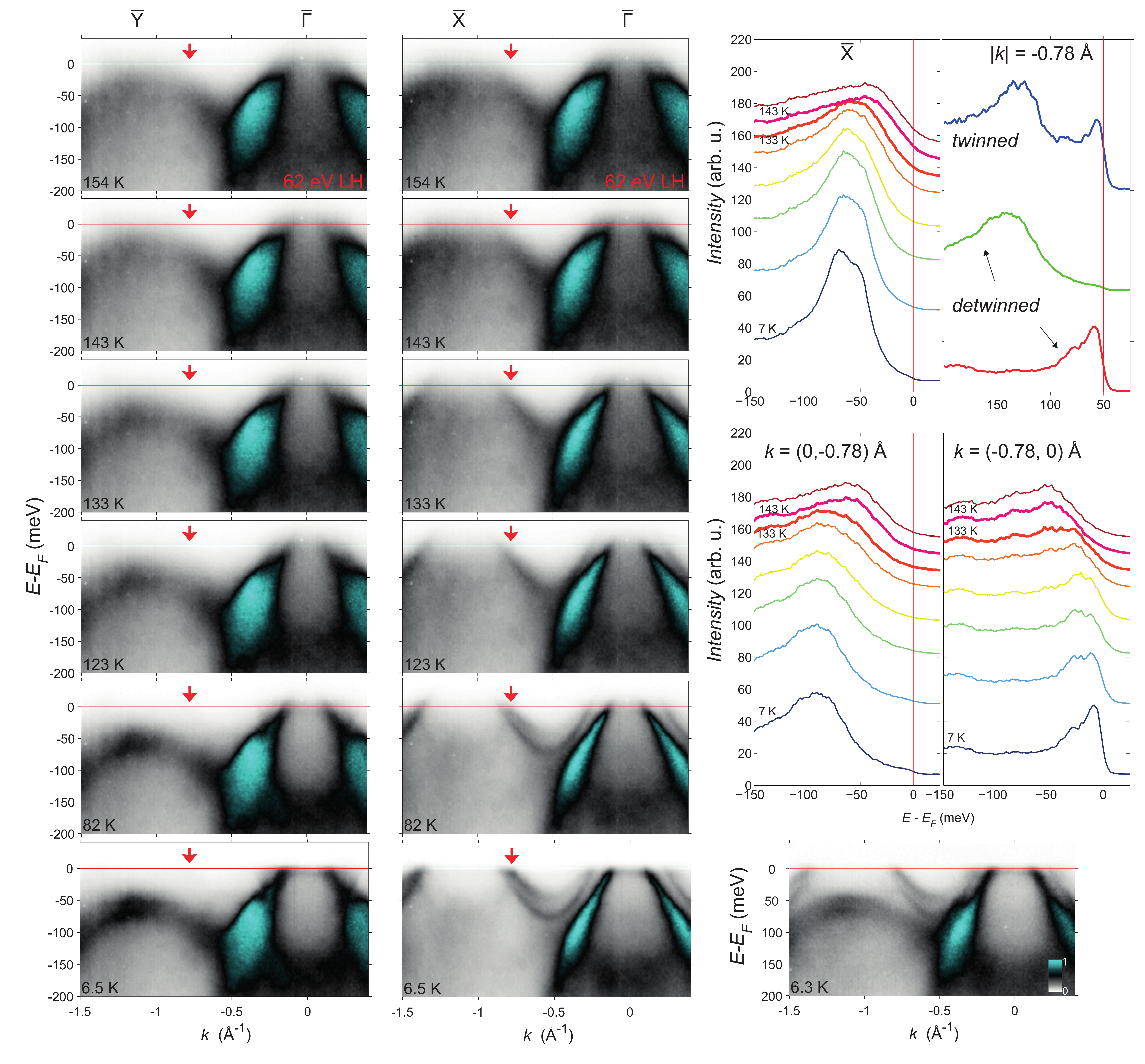}
		\caption{\textbf{Temperature-dependent measurements of detwinned BaFe$_2$As$_2$.} All data measured with 62~eV photon energy in LH polarisation; see Fig.~2 of the main text for low temperature Fermi maps taken in this geometry. The data at high temperature are significantly broadened due to much higher scattering rates at elevated temperatures. We find that there is a very little evolution of the band structure up to 123 K, but relatively abrupt changes occur between 133~K and 143~K (highlighted EDCs), i.e. below and above $T_s = T_N=$~137~K. This behaviour is consistent with the observation that the phase transition at 137~K in pure BaFe$_2$As$_2$ is first order \cite{Boehmer2015NComms}, as well as the sharp change in resistivity at this temperature \cite{Aswartham2011}.}
		\label{fig:smtdep}
	\end{figure*}
	\clearpage
	
	\section{Nano-ARPES: spectra from fixed positions}
	
	\begin{figure}[h]
		\centering
		\includegraphics[width=\linewidth]{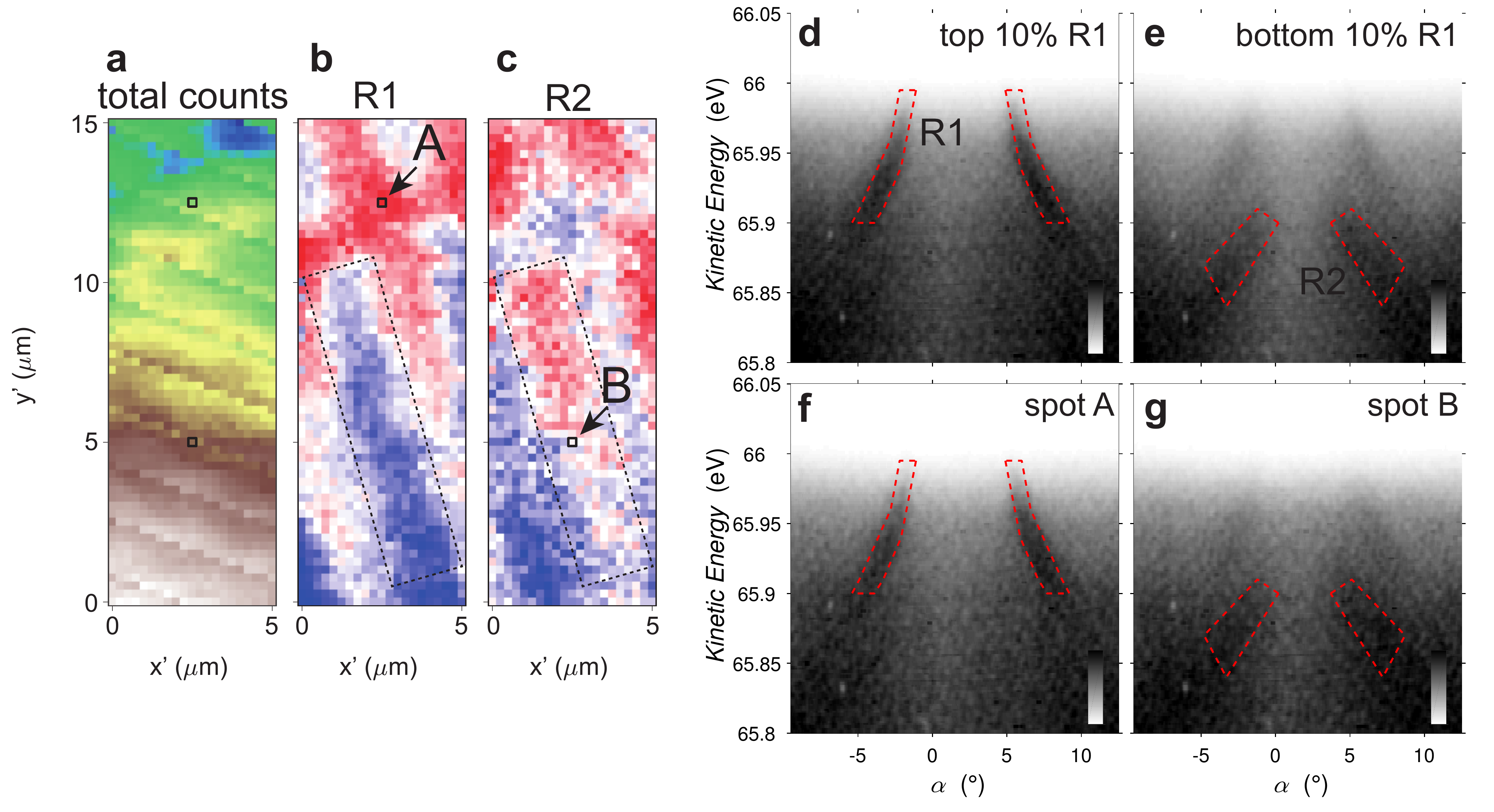}
		\caption{\textbf{Nano-ARPES spectra from within one domain.} Panels (a-e) are reproduced from Fig.~5 of the main text, except now the small boxes labelled A and B are drawn. ARPES spectra obtained with the focused beam at these positions are shown in (f-g). It can be seen that the filtering process on the spatial map, and the measurements at individual spots give qualitatively similar results.}
		\label{fig:nanospot}
	\end{figure}

\end{document}